\documentclass[11pt,oneline,nonumbib,a4paper]{ouparticle} 
\usepackage[english,french]{babel}
\usepackage{graphicx}
\usepackage[round]{natbib}

\usepackage{endfloat}
\usepackage{lineno}
\usepackage{url}

\begin{document}


\title{Enhancing and comparing methods \\
for the detection of fishing activity \\
from Vessel Monitoring System  data.}
\author{
\name{Gilles Guillot}
\address{Technical University of Denmark,\\ 
Department of Applied Mathematics and Computer Science\\
2800, Lyngby, Denmark}
\email{gigu@dtu.dk}
\and
\name{Pierre Benoit}
\address{Technical University of Denmark,\\ 
Department of Applied Mathematics and Computer Science\\
2800, Lyngby, Denmark}
\email{pierre.benoit@centraliens-lille.org}
\and
\name{Savvas Kinalis}
\address{Technical University of Denmark,\\ 
Department of Applied Mathematics and Computer Science\\
2800, Lyngby, Denmark}
\email{s131110@student.dtu.dk}
\and
\name{Fran\c cois Bastardie}
\address{Technical University of Denmark,\\ 
Department of Aquatic Resources\\
2920 Charlottenlund, Denmark}
\email{fba@aqua.dtu.dk}
\and
\name{Valerio Bartolino}
\address{Swedish University of Agricultural Sciences\\
 Department of Aquatic Resources,\\
45321 Lysekil, Sweden}
\email{valerio.bartolino@slu.se}
}
\abstract{Vessel Monitoring System (VMS) data provide information about speed and position of fishing vessels. This  
opens the door to methods of estimating and mapping fishing effort with a high level of detail. 
To addess this task, we propose a new method belonging to the class of hidden Markov models (HMM) that 
accounts for autocorrelation in time along the fishing events and 
offers a good 
trade-off between model complexity and computational efficiency. 
We carry out an objective comparison between this method and two competing approaches on a set of VMS data from Denmark for which the true activity 
is known from on-board sensors. 
The DMKMG approach proposed outperformed the competitors approach with 6\% and 15\% more accurate estimates in the vessel-by-vessel
and trip-by-trip case, respectively. In addition, these better performances are not paid in terms of computation time. 
We also showcase our method on an extensive dataset from Sweden. 
A quick (real-time) data processing has the potential to change how fisheries can be better harmonized to other utilisation of the seas 
and fill the gap between the local scales at which fishing pressure and stock depletion occurs with the large temporal 
and spatial scales of traditional fisheries assessment and management. 
The computer code developped in this work is made publicly available as an R package from \url{http://www2.imm.dtu.dk/~gigu/HMM-VMS}.
}

\date{\today}


\maketitle

\section{Background}\label{sec:intro}

The ecosystem approach to fisheries (EAF) increasingly requires spatially resolved fisheries data, and suggests the inadequacy of traditionally aggregated landing and effort data to characterize the impact of the fisheries on species, habitats and ecosystems \citep{Jennings2005,Douvere2009}. Marine spatial planning, which is recognised as an essential step towards an ecosystem-based management of the sea, demands for detailed knowledge on the spatial and temporal distribution of human activities including fishing \citep{Douvere2008}.

The implementation of a system for satellite-based geographic monitoring of fishing vessels (Vessel Monitoring System – VMS), originally for control purposes, has made possible to track and map fishing effort on a much finer scale. Over the past few years VMS data has become more widely available for scientific purposes, and it is now part of the indicators listed in the EU data collection framework \citep{EC2008}. 
This enhanced the development of new approaches and tools \citep[e.g.][]{Hintzen2012} to analyse this type of data, and the establishment of international cooperation to facilitate regional and multinational analysis of the distribution and intensity of the total fishing effort \citep[][]{ICES2014,STECF2013}.
Most of the methodological work on VMS data has focused on developing methods for dealing with inherent limitations of this data source (Lee et al. 2010), given the original purpose of control when the system was enforced in the European Union in  January 2000. Main limitations include a partial coverage of the fishing activities as VMS was initially adopted on vessels longer than 24 meters and progressively extended to smaller vessels to comprise all units larger than 12 meters since 2012. From 1 January 2005 the recording frequency must be lower than two hours but it is rarely lower than one hour. Moreover, VMS returns vessel positions regardless the activity performed by the vessel at that particular time which may comprise steaming to and across fishing grounds, searching for the catch, actual fishing and handling the catch.

Identification of behavioural states and inference on the activity of individual vessels from discrete and semi-regular observations is a challenging issue. This is particularly true considering that spatio-temporal dynamics of fishing effort are the result of the development of fishing tactics and strategies of individual fishers which are highly context- and fishery-dependent \citep{Salas2004}. Fishers show an adaptive response to changes in resource abundance and distribution, environmental conditions and market or regulatory constraints \citep{Bastardie2015} which may be potentially detected by changes in the VMS patterns.
Given the impact that misclassification of vessel activity may have on the estimation of fishing effort, the need for rigorous and reliable methods for detection of fishing activity has attracted a large amount of methodological research on VMS data \citep{Lee2010,Vermard2010}.
Recent advances in the statistical analysis of movement and behavioural data have opened opportunities for the development of generic and more rigorous statistical approaches to the analysis of VMS data. Improvements in distinguishing between different activities or states from VMS data would have multiple benefits. It would improve estimation of the effective fishing effort which can then be translated into a measure of fishing pressure on species and a measure of impact on habitats. Moreover, the different activities that characterise a fishing trip have usually different associated costs and revenues, and the amount of time spent in each has to be identified for an economic evaluation \citep{Pelletier2009,Bastardie2015}.

\begin{figure}
\begin{center}
\begin{tabular}{c}
\includegraphics[width=7.5cm]{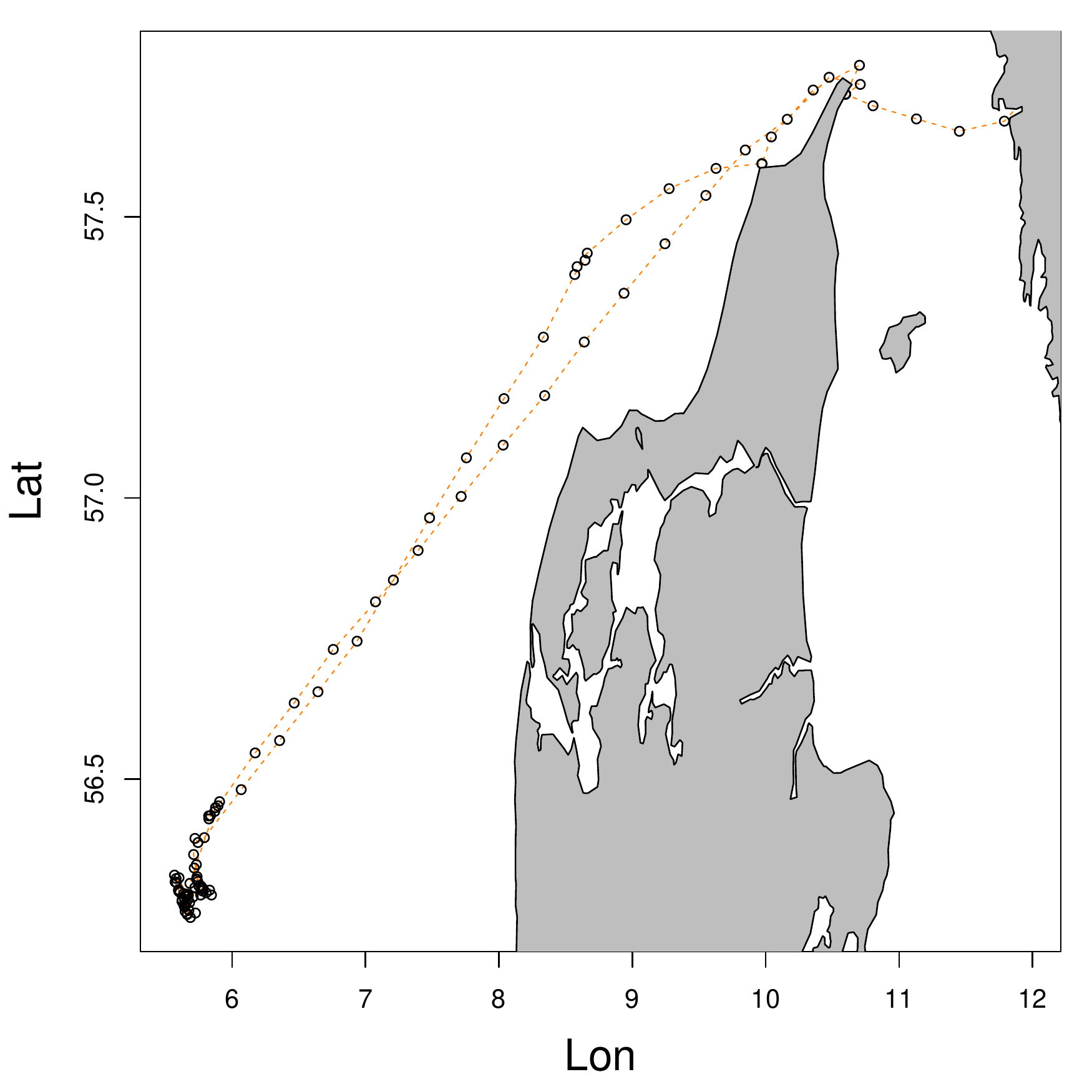}\\
\includegraphics[width=7.5cm]{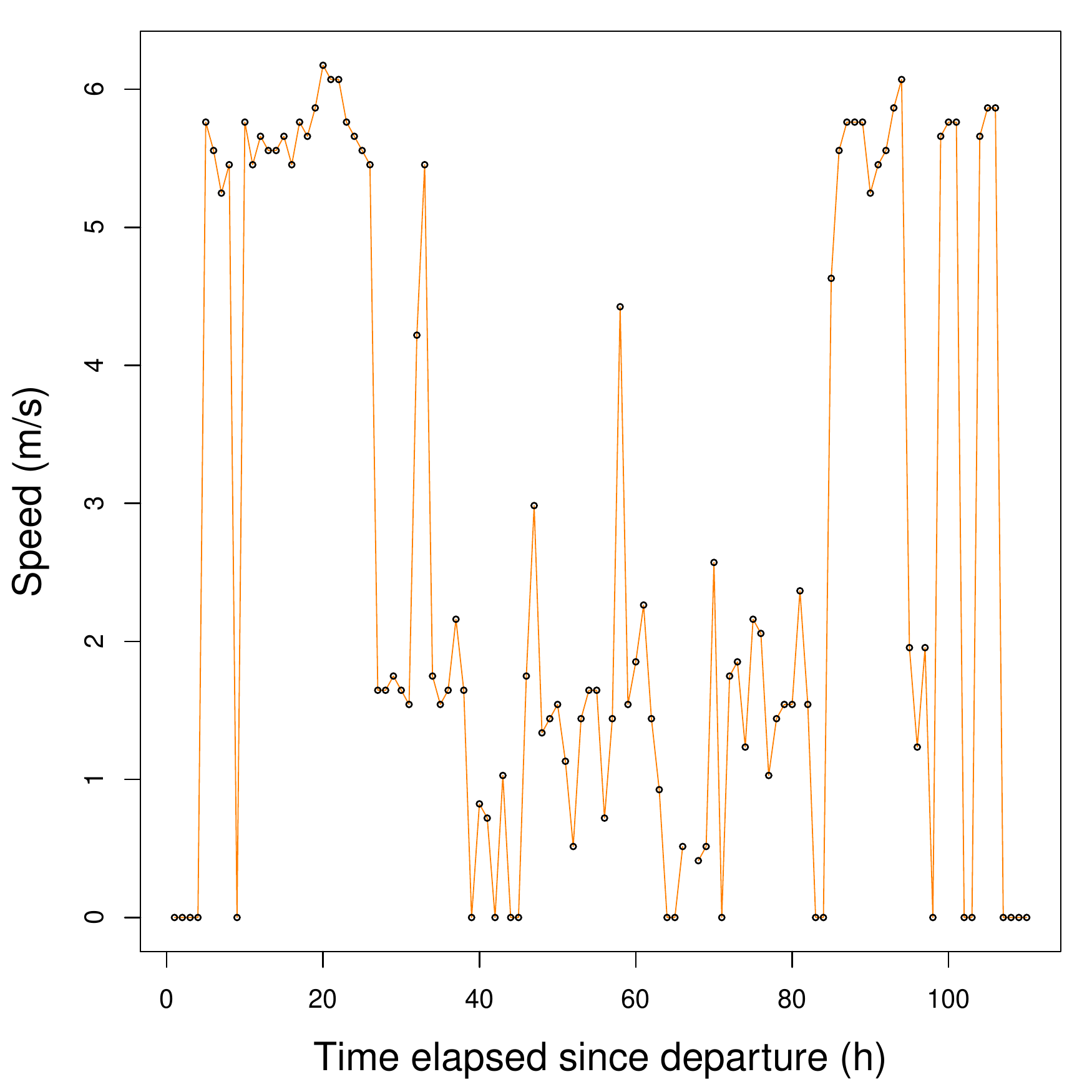}\\
\includegraphics[width=7.5cm]{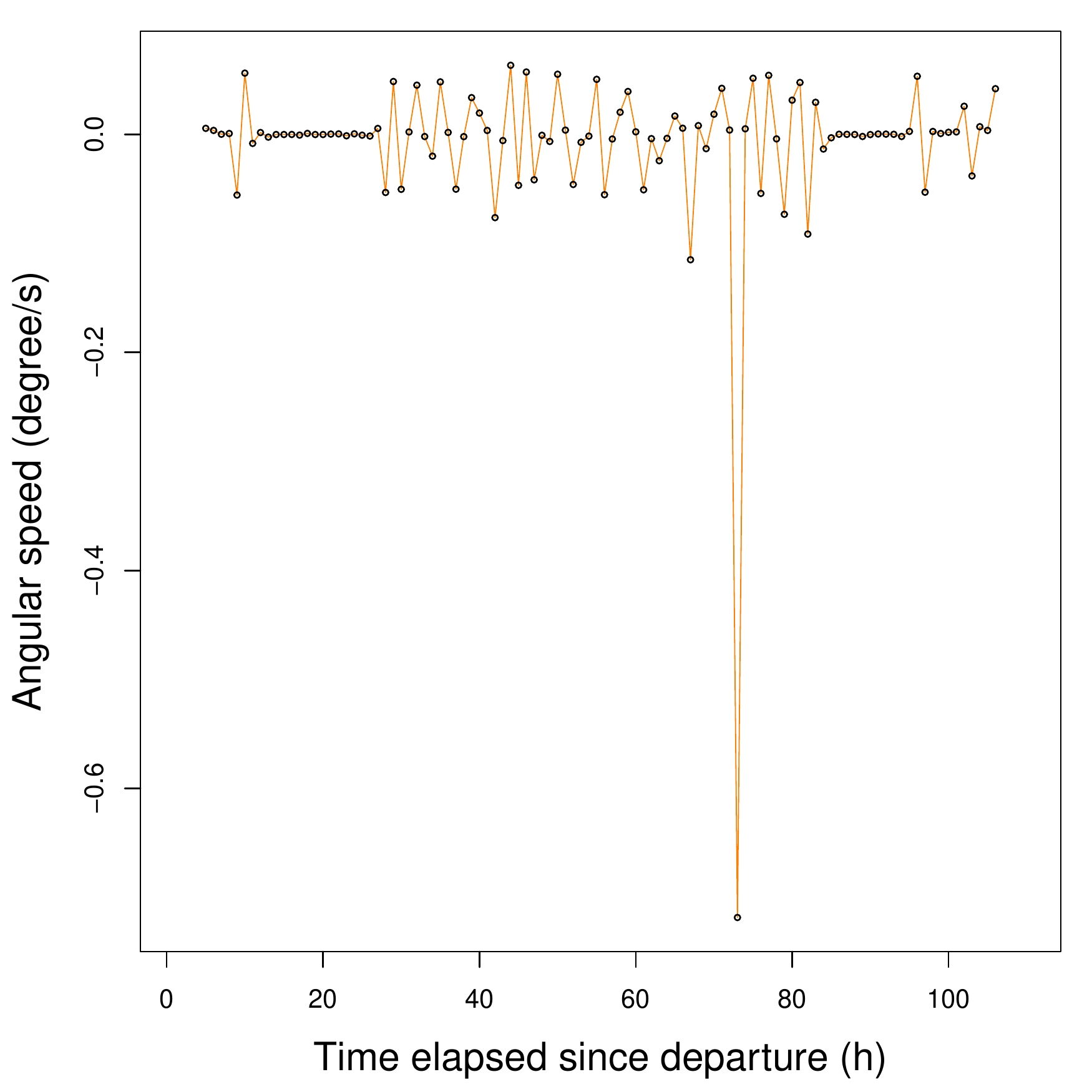}
\end{tabular}
\end{center}
\caption{Example of vessel trip trajectory}\label{fig:example_trip}
\end{figure}

The common rationale behind methods of detecting fishing activity from
Vessel Monitoring System (VMS) data lies in the fact that fishing activity results typically in slower and more erratic trajectories than steaming activity. This is clearly visible on the vesssel trip shown on Fig.~\ref{fig:example_trip} and for which brief steaming periods at high speed between the harbor and a trawling area flank a long period of erratic movements at low speed. 
A simple method to estimate the state of a vessel consists in setting
a lower and an upper threshold on the speed modulus and estimate  any ping or time step
as fishing if it is within these thresholds. This method is implemented for example 
in the VMStools program \citep{Hintzen2012} and became   
 a widely used approach across Europe for processing VMS and logbook fisheries data
  \citep[see also][for review and
further references]{Lee2010}. 
A potential weakness of this method it that disregards any auto-correlation in times, in particular, 
the facts  that a vessel does not alternate between  fishing and steaming 
states too frequently. In principle, this simple threshold method can be improved in many ways. 
\citet{Mills2007} proposed to combine both the speed modulus and directionality in such a threhsold-based method. 
\citet{Bastardie2010} proposed to account for auto-correlation in time of vessel states by using a hidden Markov model (HMM).
\citet{Peel2011} enhanced the previous method by accounting for other states than fishing and steaming such as  exiting, entering 
and staying at an anchorage. They also take into account the time of day in the estimation. 
\citet{Gloaguen2014} proposed to model auto-correlation in time of the vessel speed within each state with an HMM model. 
This contrasts with the methods of \citet{Bastardie2010} that assumes that speeds within a fishing segment or within a steaming segment are independently distributed. 
Both \citet{Peel2011} and \citet{Gloaguen2014}  harness parameter estimation to maximum likelihood computation and 
the EM algorithm.
We note also that earlier works differ in terms of type of validation carried out,  
\citet{Mills2007} and \citet{Bastardie2010} confronting their estimates to the known activity available 
from on-board camera or observers data, while \citet{Peel2011} and \citet{Gloaguen2014} restricting their investigation to the use of simulated data. 

The premise of the present paper is that model-based clustering methods and in particular hidden Markov models offer a flexible yet computationally efficient framework for the estimation of vessel activity and deserve further attention. 
The work is developed in three steps. First, we investigate in detail the application of a new HMM model and several variants in their practical implementation. Second, we compared the new approach to a number of existing methods, including the threshold-based method implemented in VMStools and to the model-based method 
of \citet{Gloaguen2014}. To do so, we evaluate the accuracy of methods with the aid of on-board sensor validation data. Third, we implement the model found to be the most accurate on an extensive dataset from the Swedish bottom trawl fisheries operating in the Baltic and the Kattegat-Skagerrak. The Swedish bottom trawlers target a number of key demersal species in these areas and contribute to more than 20\% of the total EU catch of cod (\textit{Gadus morhua}) in the Baltic and 
more than 25\% of Norway lobster (\textit{Nephrops norvegicus}) catch in the Kattegat-Skagerrak area.

\section{Material and methods}

\subsection{Data}
To design our model and showcase its main features, we study here a
set of VMS data consisting of 6430 fishing trips carried out in 2009 by 131 Swedish bottom trawlers operating in the Baltic (ICES subdivision 24-25) and the Kattegat-Skagerrak (ICES subdivision 20-21). VMS data are available from the Swedish Agency for Marine and Water Management with link to fishermen logbooks based on individual vessel signal and date-time information on the trip. According to information from the logbooks, VMS data associated to specific fleet segments (metier unit at the DCF level 5) were selected for particular combinations of gear type (i.e., otter-trawlers) and target species assemblage (i.e., demersal fish and crustaceans) as calculated from the reported landings.

In order to assess the accuracy of the methods studied here, we used a validation dataset of
ten trips made by six Danish vessels and for which the actual state (fishing or steaming) is
exactly known from sensors attached to the fishing gear \citep{Bastardie2010}. 
Both VMS datasets feature approximately a one-hour ping frequency.

\subsection{Model proposed for the clustering of VMS data} \label{sec:our_method}

We denote by $v_t$ the linear speed (speed modulus in the direction of
movement) and $\omega_t$ the angular speed (change of direction per
time unit) of the vessel at time $t$. 
These variables can be either available directly from on-board sensors
as part of the VMS monitoring system or derived numerically as finite
differences from GPS positions available readily as VMS data. 
Following a standard approach in model-based clustering
\citep{Mclachlan00}, we introduce a discrete variable $s_t$ defined at each time step and encoding the unknown state of the vessel. 
In the simplest instance, $s_t$ has two states coded as $1/2$, standing for fishing and steaming, and 
we assume that conditionally on $s_t=k$ ($k$=1 or $2$), the vector $(v_t,\omega_t)$ arises from a  bivariate Gaussian distribution with 
state-specific mean $\mu_k$ and variance matrix $\Sigma_k$. 
The difference between $\mu_1$ and $\mu_2$ reflects the difference of
sailing behavior during fishing and steaming. 
Because the data we analyze are not perfectly fitted by a two-cluster Gaussian mixture 
(cf. Fig.~\ref{fig:histo_speed_Dk}, left panel), 
we consider a  class of models in which the number of states $K$ can be larger than $2$. 
For example, in a model where $K=3$, a Gaussian component will typically fit
fishing states while the two other components will fit the various
steaming states, one for the high-speed steaming states and the other 
one for the various low-speed steaming states (cf. Fig.~\ref{fig:histo_speed_Dk}, right panel). 

As pointed out earlier and illustrated by figure \ref{fig:example_trip},
a vessel does not alternate constantly between fishing and steaming
states. Using a statistical phrasing, the sequence of states
$(s_1,...,s_T)$ is not properly model by a sequence of independent
variables. 
Injecting this information in the model is presumably a good way to
increase accuracy of inference. We do this by mean of a Markov chain 
in which the probability of the next state depends on the current
state. The set of conditional probabilities
$p_{i,j}=p(s_{t+1}=j|s_t=i)$ forms 
a matrix known as transition matrix \citep{Ross2007} and is part of the
unknown quantities to be estimated. The model formed by the Gaussian
components together with the Markov chain for the un-observed states belongs to the
class of hidden Markov models  (HMM) which is an
important tool in science and engineering \citep{Zucchini2009} to model heterogenous data, in particular multi-modal data. The
particular models we consider here can be referred to as dependent
mixture of K multivariate Gaussian models (DMKMG). 
To estimate parameters, we use likelihood maximization for HMMs by the EM algorithm 
 implemented in the  R package  {\tt depmixS4} \citep{Visser2010}.  \\

In a model with $K=2$ components, the component estimated with the lowest mean speed 
has to be interpreted as the component modeling Fishing states 
(cf. Fig.~\ref{fig:histo_speed_Dk} for an illustration). 
As soon as we make use of a model that includes more than two components, 
the interpretation of the various components is not straightfoward. 
A possible approach consists in imposing some constraints on the parameters at the estimation stage, 
for example constraining the 
mean speed of a certain component (that would stand for Fishing states) to be e.g. less than 3 knots. 
Pilot experiments along this line led to convergence issues with the EM algorithm. 
Besides, constraining parameters does not fully address the issue of the 
interpretation of the components. Therefore we did not pursue this approach. 
We propose instead an alternative strategy in which the parameters are not constrained during the parameter estimation stage 
and the labelling is  performed after parameter estimation. 
It is based on the observation that the low-speed component for a 2-component model has a variance that is too 
large because it includes many Steaming states.
Our labelling algorithm goes as follows: we label as Fishing the combination of components 
that has a joint empirical variance smaller than that of the low-speed states estimated in a 2-component model. 
If there is more than one combination of components that reduces the variance, we choose the one with a 
joint empircal mean closest to that of the low-speed component estimateed in a 2-component model.

\subsection{Competing methods considered here }
We compare outputs of the method outlined above to those obtained
first with a simple threshold-based method and implemented in the
program {\tt VMStools} \citep{Hintzen2012} and second with the method proposed by
\citet{Gloaguen2014}. 
In the model of \citet{Gloaguen2014}, the speed is parameterized
in terms of a persistence speed (along the direction at the current
time step)  and a rotational speed (along the direction at the next
time step). This model assumes  the existence of two states (Fishing,
Steaming) whose distribution is given by a Markov chain. 
Because it is reasonable to assume that within a state, the speed
varies smoothly, \citet{Gloaguen2014} place a time dependence structure
on each component of the speed vector in form of a first-order
auto-regressive Gaussian process. Their model appears therefore as
a time-dependent mixture of auto-regressive processes (DMARP). They carry out
inference with the EM algorithm. In the sequel, we use the R code 
developed by these authors (P. Gloaguen, personnal communication).  
A comparison to the model proposed by \citet{Peel2011} would have been relevant here but 
the code developped by these authors has not been made available.

\subsection{Validation against fine resolved data}
The model we outlined in section \ref{sec:our_method} can be
implemented with any number  $K$ of Gaussian components. It can be run either
trip-by-trip (therefore with trip-specific parameters), or vessel-by-vessel, i.e. 
treating all trips of a vessel as a single long trip (therefore with vessel-specific parameters) or even 
all data as one trip, i.e. treating all the data as a single long trip (therefore with a common set of parameters for the whole dataset). 
Besides, the model can be implemented using both linear and angular
speed or linear speed only (assuming the use of angular speed only
would not make sense). This defines up to K$\times$3$\times$2 sub-models. 
The first goal of the present analysis is to assess which sub-model 
yields the most accurate results. 
To do so, we run each sub-model on the validation dataset, obtain a 
vector of estimated fishing states and compare it to the true state
available from on-board camera data. The congruence between the vector
of true and estimated states is assessed numerically through four
statistics:
\begin{equation}
\mbox{Global match} = \frac{\#\{\mbox{Estimated state} == \mbox{True
  state}\}}{\# \{ \mbox{Time steps} \}}
\end{equation}
\begin{equation}
\mbox{F as S} =  \frac{\#\{\mbox{Fishing state estimated as Steaming state}\}}{\# \{ \mbox{Time step} \}}
\end{equation}
\begin{equation}
\mbox{S as F} =  \frac{\#\{\mbox{Steaming state estimated as Fishing state}\}}{\# \{ \mbox{Time step} \}}
\end{equation}
\begin{equation}
\mbox{Unestimated state} =  \frac{\#\{\mbox{Missing estimated state}\}}{\# \{ \mbox{Time step} \}}
\label{eq:Unest_state}
\end{equation}
In equation \ref{eq:Unest_state}, missingness corresponds to states for which
inference failed at least partly, typically for short trips that
bring up indentifiability issues. 
In a first set of pilot computations, we compared accuracies obtained
with both linear and angular speeds to those obtained with linear
speeds only. We carried out comparisons trip-by-trip, vessel-by-vessel and all
data together for $K=2$. 
In a second set of analyses, we estimated vessels states with the DMKMG
model using linear speeds alone with a number of
Gaussian components K ranging from 2 to 10. Again we carried out comparisons trip-by-trip, vessel-by-vessel and all
data together. 
In a third set of analyses we compared the accuracy of the sub-model we
found to be the most accurate in the analysis described above, 
to that obtained with VMStools and the DMARP
model \citep{Gloaguen2014} described in the previous section.

\subsection{Application to Swedish data} 
We use the model found to be the most accurate on the Danish validation data to map the fishing effort 
from the Swedish demersal trawl fleet in 2009.
In a vector of estimated states $\hat{s}_1,...,\hat{s}_T$ we consider that a stretch of consecutive states estimated as 
Fishing states form an estimate of a trawling event. 
Once the fishing states are separated from the other states, we map the fishing effort for this period over the western Baltic Sea and the Kattegat-Skagerrak. 

\section{Results}

\subsection{Danish data}
\begin{figure}[h]
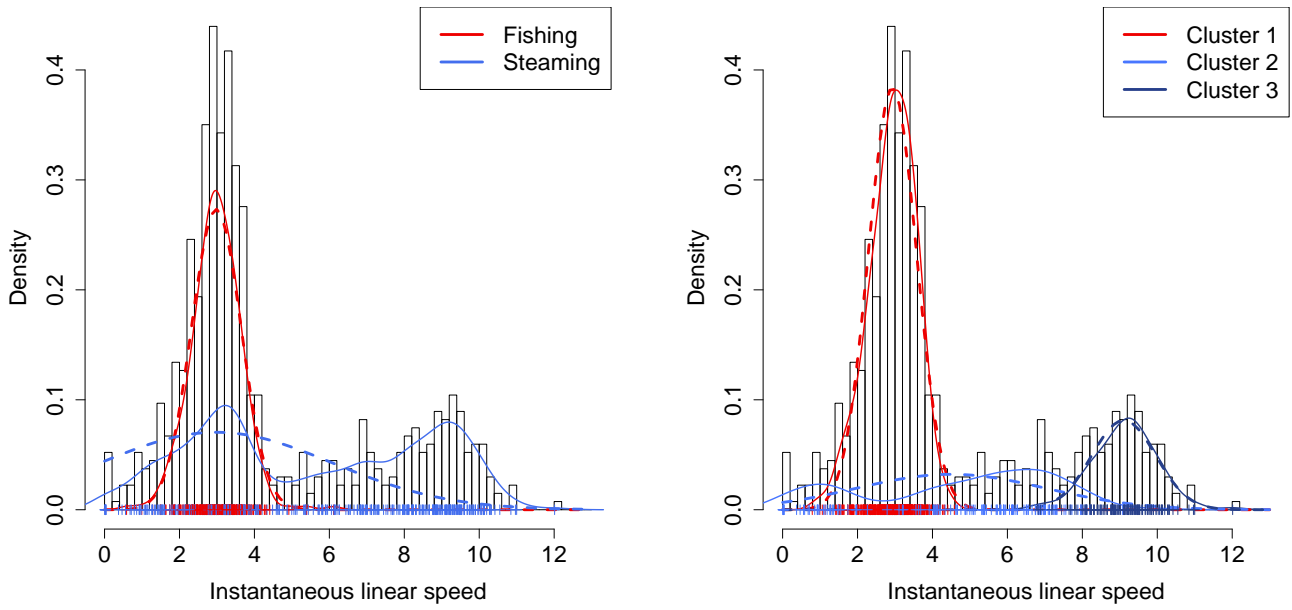

\hspace{-.03cm}\begin{tabular}{cc}
\hspace{-1cm}\includegraphics[width=8.5cm]{./hist_dens_true_state.png} & 
\includegraphics[width=8.5cm]{./hist_dens_est-depmix_3clusters.png}
\end{tabular}
\caption{Histogram of speed data on the Danish dataset. 
Left: true state, right: estimated state with a 3-component dependent Gaussian mixture. 
Continuous line: smoothing of empirical histogram (non-parametric density), dashed line: theoretical Gaussian component fitted. 
}\label{fig:histo_speed_Dk}
\end{figure}

Under the DMKMG model, we observed that making use of
angular speeds together with linear speeds does not bring any
improvement, or even lead to lower accuracies, than estimates 
using linear speed only, all other options being equal. 
Using linear speeds only, the DMKMG model yielding the highest accuracy in terms of global match
is a model with $K=3$ Gaussian components. 
Figure~\ref{fig:histo_speed_Dk}
illustrates the fit of this model and the empirical and theorethical distributions of known steaming and fishing states from the validation dataset. Results show that an extra Gaussian component helps capturing some of the low speed steaming events. 
Note that the overlap of red and blue dots on the x-axis in the right panel is due to the use of a HMM that does not classify states 
according to the estimated density value \cite[cf.][]{Zucchini2009}.

Accuracies of this best model together with those obtained with the
competing methods are reported in table~\ref{tab:results_Dk}. 
In the sequel, unless specified otherwise, all results refer to this best model.


The DMKMG model (three Gaussian components, linear speed only) shows the highest level of global match followed by the threshold method implemented by VMStools and then by the DMARP model. In the 'all data together' comparison, both DMKMG and VMStools are able to estimate all the states and achieve a correct identification score of 81.5\% and 79.0\%, respectively. DMARP achieves only a 68.5\% global match due to 1.5\% of unestimated states and mostly because of approximately 30\% of steaming as fishing mis-classicifations compared to the 17.5\% of DMKMG and 19.6\% of VMStools. In the vessel-by-vessel and trip-by-trip comparison DMKMG maintains same high levels of global match (i.e., 78.6\% and 79.9\%, respectively). On the contrary, VMStools shows a decrease in performances (i.e., 72.9\% and 64.5\%, respectively), comparable to the matching level achieved by DMARP, due to an increased number of unestimated states.

\begin{table}[h]
\caption{Accuracy in estimating vessels activity on Danish data. All
  values are percentages. Number in parentheses are obtained when counting unestimated states as errors. 
See text for detail. }\label{tab:results_Dk}
\vspace{1em}
\begin{tabular}{llccc}
\hline
\hline
\vspace{.5em}
Method & Error statistics & \begin{tabular}{c}all data\\ together \end{tabular}& vessel-by-vessel & trip-by-trip \\[.5em]
\cline{2-5}
VMStools & 
\begin{tabular}{l}  Global match \\    F as S  \\ S as F \\ Unestimated  states\end{tabular}& 
\begin{tabular}{c} 78.99        \\ 1.34        \\ 19.67 \\ 0 \end{tabular}&
\begin{tabular}{c} 77.61 (72.88)\\ 1.59 (1.49) \\ 20.79 (19.52) \\ 6.11 \end{tabular}&
\begin{tabular}{c} 78.48 (64.48)\\ 1.62 (1.34) \\ 19.89 (16.39) \\ 17.58 \end{tabular} \\[2.5em]
\cline{2-5}
DMARP & 
\begin{tabular}{l} Global match \\ F as S     \\  S as F \\ Unestimated states \end{tabular}& 
\begin{tabular}{c} 69.6 (68.55) \\ 0.3 (0.3)   \\ 30.10 (29.66)\\ 1.51 \end{tabular}&
\begin{tabular}{c} 70 (69.00)   \\ 3.47 (3.43) \\ 26.48 (26.08) \\1.51\end{tabular}&
\begin{tabular}{c} 66.57 (65.57)\\ 2.11 (2.09) \\ 31.32 (30.85) \\ 1.51 \end{tabular}  \\[2.5em]
\cline{2-5}
\begin{tabular}{l}Best DMKMG \\ model\end{tabular}& 
\begin{tabular}{l} Global match \\ F as S \\ S as F \\ Unestimated states \end{tabular}& 
\begin{tabular}{c} 81.52 (81.52)\\  0.94 (0.94) \\ 17.54 (17.54) \\ 0 \end{tabular}&
\begin{tabular}{c} 78.87 (78.62)\\  1.30 (1.30) \\ 19.82 (19.76) \\ 0.32 \end{tabular}&
\begin{tabular}{c} 80.25 (79.94)\\  2.10 (2.09) \\ 17.65 (17.58) \\ 0.39 \end{tabular}  \\[2.5em]
\cline{2-5}
\end{tabular}
\end{table}

\clearpage
\subsection{Swedish data}
\begin{figure}[h]
\hspace{-2cm}
\begin{tabular}{c}
\includegraphics[width=19cm]{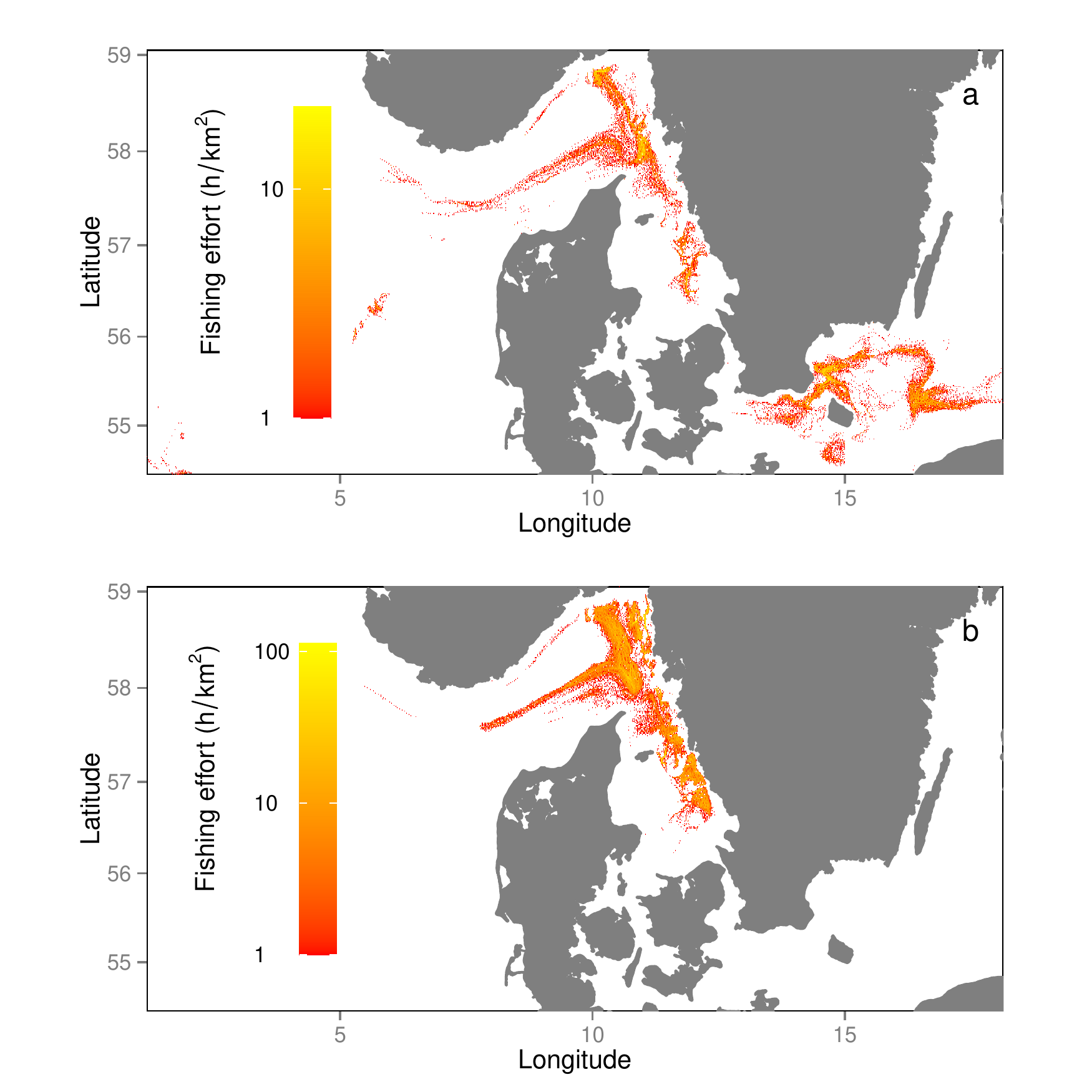}
\end{tabular}
\caption{
Swedish fishing effort for the demersal fish (upper panel) and crustaceans (lower panel) bottom trawlers (graduation on the log-scale).
}\label{fig:fishing_effort}
\end{figure}

Figure~\ref{fig:fishing_effort} represents the spatial distribution of the fishing effort of the Swedish bottom trawl fisheries in the western Baltic (ICES subdivision 24-25) and the Kattegat-Skagerrak area (ICES subdivision 20-21) in 2009. In the western Baltic, the Swedish fisheries is highly concentrated in the waters northen Bornholm Island between the subdivisions 24 and 25, and along the 50-70 m isobaths that stretch in the northern and eastern side of the Bornholm Basin. The central and deepest part of the Bornholm Basin is mostly free from bottom trawling effort. Fishing effort is widely distributed along the Swedish Kattegat and Skagerrak on soft bottoms in the range of 50-200 m depth. Small and medium size un-trawled patches are visible throughout the area. In the Skagerrak the fishing effort of the bottom trawl fisheries extends westward into the North Sea, bounding the southern side of the Skagerrak deep ($>$300 m depth) which is left free from bottom trawling. 



\clearpage
\section{Summary and discussion}

We have implemented various methods to detect fishing activity from VMS data. 
The method that performs best on the Danish data is the dependent mixture with three components. 
Presumably because it achieves the best  trade-off between model parsimony and complexity. 
On the Danish trawlers data, the classification error can be brought below 20\% and the proportion of Fishing states estimated as Steaming around 1\%. We note that these numbers are likely lower on the Swedish data as the vessels are similar 
but parameters estimation is performed on a much larger dataset.
In supplementary material, we report further investigation about the poor performances obtained 
when using information about the change of direction of the vessel. 
It is seen that in the various parametrisations of the speed vector, information carried by 
directional data can not be used efficiently by Gaussian mixture models. \\
One of the question that triggered this work was whether logbook data could be used to increase 
accuracy of  algorithm for the detection of vessel states. Here we observed that errors of methods based on 
VMS data only correspond to isolated low-speed steaming states. It seems therefore unlikely that information about trawling fishing activity present in the logbook data could prevent this type of errors. \\

\cite{Peel2011} suggest, and this is confirmed by a preliminary analysis of our validation
data, that fishing  events tend to occur preferentially at a specific time of the day
(day time in our Danish data). 
It would therefore be natural to inject this property in our model by 
linking the vector of transition probabilities $(p_{i,j})_{j=1,...,K}$ to the time of
the day $h_t$ using a multinomial logistic model \citep{Visser2010}. 
However, the time of the day is a circular variable 
($h_t=$ 0:00 is equivalent to $h_t=$ 24:00) and this is not well handled by the link functions
implemented in {\tt depmixS4}. This seems to be a promising direction to further improve accuracy 
of fishing detection methods in this context. 

The computer code we used for the K-component Gaussian mixture model is based on the R package {\tt depmixS4} programmed
 in C.  
This allows us to obtain computing time of the order of ten minutes to carry out inference on the whole Swedish dataset (131 vessels, 6430 trips)
on a standard single-processor computer. This opens the door to accurate real-time information about accumulated fishing effort 
for decision makers \citep{Kraak2012,Needle2011}. 
The increasing availability and coverage of high-resolution real-time data on vessels positions (i.e., VMS and Automatic Identification System) and catches (i.e., electronic logbook) is profoundly changing the way we look at fishing pressure and impact on the marine systems. This has the potential to change how fisheries can be managed in the future and better harmonized to other utilisation of the seas (the maritime spatial planning is central in the development of the MSFD as stated in the EU MSP 2014/89/EU directive).
One of the main cutting edge use of such high resolution data consists in the potential reconciliation of the existing gap between the small and local scales at which fishing pressure and stock depletion may occur and the large temporal and spatial scales of traditional fisheries assessment and management \citep{Bartolino_etal2012}. Recently, new ideas for real-time management of fisheries have been proposed \citep{Holmes_etal2011,Needle2011}. For instance, the real-time incentive (RTI) system is stimulating a discussion on alternative forms of fisheries management which may raise from application of these new technologies \citep{Kraak2012}. Among the different aspects at discussion there are issues concerning the potential response of fisheries and occurrence of unpredicted and unlikely behaviour in fishermen, uncertainty in the assessment of the resources and value of habitat features impacted by the fisheries, technical requirements involved in the monitoring and efficient real-time processing of extensive information on the fisheries behaviour and catch \citep{Kraak_etal2014}. 
The DMKMG approach proposed showed a marginal improvement compared to the other methods in the 'all data together' conditions investigated, but it outperformed the competitors with 6\% and 15\% more accurate estimates in the vessel-by-vessel and trip-by-trip case, respectively. In addition, these better performances are not paid in terms of computation time which as the EM resorting to the {\tt depmixS4} package reduce computing time by one or two order of magnitude, 
moving forward the current limits of real-time computation of VMS for fisheries management.

\subsection*{Funding}
This work is funded by the Swedish Research Council FORMAS under the Research and Development Project Grant 2012-94.

\bibliographystyle{plainnat}
\bibliography{biblio}


\end{document}